\documentclass[a4paper,prd,twocolumn,showpacs,superscriptaddress,floatfix]{revtex4-1}

\usepackage{graphicx}
\usepackage{dcolumn}
\usepackage{bm}
\usepackage{hyperref}
\hypersetup{
  colorlinks=true,        
  linkcolor=blue,         
  citecolor=cyan,         
  urlcolor=cyan           
}
\usepackage{colortbl}
\usepackage{amsfonts,amsmath}
\usepackage{amssymb}
\usepackage{times}
\usepackage{times}
\usepackage{multirow}
\usepackage[normalem]{ulem}
\newcommand{\bwt}{\begin{widetext}}
\newcommand{\ewt}{\end{widetext}}
\newcommand{\beq}{\begin{equation}}
\newcommand{\eeq}{\end{equation}}
\newcommand{\bea}{\begin{eqnarray}}
\newcommand{\eea}{\end{eqnarray}}

\begin{document} 
\title{Bayesian inference of signatures of hyperons inside neutron stars}
\author{Tuhin Malik}
\affiliation{CFisUC, Department of Physics, University of Coimbra, PT 3004-516 Coimbra, Portugal}
\email{tm@uc.pt}

\author{Constan\c ca Provid\^encia}
\affiliation{CFisUC, Department of Physics, University of Coimbra, PT 3004-516 Coimbra, Portugal}
\email{cp@uc.pt}
 
\begin{abstract}
The possible signatures of the presence of hyperons inside neutron stars are discussed within a Bayesian inference framework applied to a set of models based on a density dependent relativistic mean field description of hadronic matter. Nuclear matter properties, hypernuclei properties and observational information are used to constraint the models. General properties of neutron stars such as the maximum mass,  radius, tidal deformability, proton fraction, hyperon fraction and speed of sound are discussed. It is shown that the two solar mass constraint imposes that neutron stars  described by equations of state that include hyperons have in average a larger radius, $\gtrsim 0.5$km,  and a larger tidal deformability, $\gtrsim 150$, than the stars determined from a nucleonic equation of state, while the speed of sound at the center of the star is more than 25\% smaller. If a 1.4 M$_\odot$ star with a radius $\lesssim 12.5$ km is measured it is quite improbable that a massive star described by the same model contains hyperons. A similar conclusion is drawn if a two solar mass star with a radius $\lesssim 11.5$ km   or a neutron star with a mass above 2.2 M$_\odot$ is observed: the possible hyperon content of these stars is ruled out  or  very reduced. The hyperon presence inside neutron stars is compatible with the present NICER mass-radius observations of the pulsars PSR J0030+0451 and PSR J0740+6620 and the gravitational wave detection GW170817.
{It is shown that if the polytropic index $\gamma=\partial\ln p/\partial\ln\epsilon$  takes values of the order of 1.75 at not too large densities, it may indicate the onset of some kind of exotic matter, but not necessarily  of deconfined quark matter.}
\end{abstract}


\maketitle

\section{Introduction}
{Recent developments in multi-messenger astronomy bring important information on matter far beyond the one reachable in terrestrial laboratories, e.g., the very large neutron-proton asymmetry and baryonic density. The core of neutron stars (NS) may contain extreme phases of matter \cite{book.Glendenning1996}. The NS observed as pulsars are one of the densest and  most compact objects in the universe.} In {the year} 2010, when the massive pulsar PSR~J1614-2230, the first high mass neutron star (NS) with a mass close to two solar masses, was announced with a small uncertainty at 68\% confidence interval (CI), the authors have questioned whether the presence of exotic degrees of freedom such as hyperons, pion or kaon condensates or a deconfined quark phase would be compatible with such a large mass  \cite{Demorest2010}. This was due to the fact that the onset of new degrees of freedom inside NS would soften strongly the equation of state, not allowing that stellar matter would be able to support  masses as large as a two solar masses \cite{Baldo:1999rq,Vidana:2000ew,Schulze:2006vw}. Later, the mass of the PSR~J1614-2230 was reviewed \cite{Fonseca2016,Arzoumanian2017}, being the present accepted value $M = 1.908 \pm~ 0.016 M_{\odot}$. Meanwhile, the mass of  other two  solar mass NS have been determined: PSR~J0348 - 0432 with $M = 2.01 \pm~ 0.04~ M_{\odot}$ \cite{Antoniadis2013},  PSR J0740+6620 with $M = 2.08 \pm~ 0.07~ M_{\odot}$  \cite{Fonseca:2021wxt} and, very recently, J1810+1714 with a mass $M = 2.13 \pm~ 0.04~ M_{\odot}$  \cite{Romani:2021xmb}.  Soon, after the announcement of PSR~J1614-2230, several authors have shown that two solar masses are still compatible with exotic degrees of freedom taking into account the still reduced information on the strong force at high densities, see for instance \cite{Bednarek:2011gd,Weissenborn:2011kb,Weissenborn:2011ut,Bonanno:2011ch,Providencia:2012rx}, see also \cite{Oertel:2014qza,Fortin2016,Oertel:2016xsn,Fortin:2017cvt}.

{In \cite{Fortin:2014mya}, the authors have considered three nucleonic EOS and 14 hyperonic EOS proposed by different groups and analyzed the differences between the NS properties of these two sets of models. They concluded that the second set, which included hyperons, predicted larger radii, in particular a gap larger than 1 km for the radius of 1.4$M_\odot$ stars. However, they also pointed out that the 14 hyperonic EOS were obtained from models that did not satisfy the constraints obtained  for   pure neutron matter (PNM) using {\it ab-initio} methods within a chiral effective field theory framework ($\chi$EFT)  \cite{Hebeler2013}. This approach allows a controlled estimate of the uncertainties on the PNM EOS, see for a recent review \cite{Drischler:2021kxf}.  Other works have studied the effect of hyperons in stellar matter. { In \cite{Traversi:2020aaa} the authors have used a Bayesian inference approach to constrain a RMF model with non-linear meson terms and introduced the $\Lambda$-hyperon.  They have tested different types of priors for the nuclear matter properties and have predicted that a 1.4$M_\odot$ star has a radius of $\approx$12 km  if no hyperons are included in the EOS and $\approx$14 km if matter contains the $\Lambda$-hyperon. This result agrees with the conclusion of \cite{Fortin:2014mya} but predicts larger radii for 1.4$M_\odot$ stars.} {In this study, the authors have imposed only observational data within a Bayesian inference framework to constrain their RMF model parameters and have not considered any type of low density constraint such as the PNM EOS from $\chi$EFT. A different study, also within a Bayesian inference approach,  has been performed in \cite{Sun:2022yor}, where the hyperon-meson couplings were constrained by astrophysical observations. The possible correlations between nuclear and  hypernuclear parameters and NS properties have been explored within a RMF description of hadronic matter including hyperons by imposing multi-physics constraints filters, including PNM EOS from $\chi$EFT, to a randomly generated EOS set in \cite{Ghosh:2022lam}.}

{In the following, we will investigate which signatures could indicate the presence of hyperons inside NS that satisfy not only the two solar mass NS constraints but also the PNM constraints from $\chi$EFT \cite{Hebeler2013} within a Bayesian inference framework. We will first reanalyze the conclusions drawn in \cite{Fortin:2014mya} and study the acceptable domain of hyperonic EOSs in light of the present observations.} {Notice that we are assisting to a wider and wider interest on the problem of extracting the EOS of stellar matter from the knowledge of observational data, generally using statistical methods such as Bayesian inference \cite{Ozel:2010fw,Steiner:2010fz,Steiner:2012xt,Ozel:2015fia,Raithel:2016bux,Raithel:2017ity} or non-parametric inference \cite{Landry:2018prl,Essick2019,Landry:2020vaw} or deep machine learning techniques \cite{Fujimoto:2017cdo,Fujimoto:2019hxv,Ferreira:2019bny,Morawski:2020izm}. Still another problem is the determination of the star composition once the EOS is known. 
Some recent studies have shown that the extraction of the neutron star composition is not trivial even if the problem is restricted to the proton fraction \cite{Tovar2021,Imam2021,Mondal2021,Essick2021}. The present study will add another contribution to this topic, but now considering also non-nucleonic degrees of freedom.}

{In the present work,} a minimal number of nuclear matter properties will be imposed together with the observational two solar mass constraint. {Starting from a relativistic mean field description of stellar matter, considering a field theoretical model with density dependent couplings {to the meson fields} (DDH) as introduced in \cite{Malik:2022zol}, we will include hyperonic degrees of freedom and will determine the occurrence probability  of  different nuclear matter properties (NMPs) and stellar properties, within a Bayesian inference {framework}.}  For the hyperon couplings, we will restrict ourselves as much as possible to a reduced  coupling domain. For the vector mesons we will take the predictions from the SU(6) quark model, and the coupling to the $\sigma$  field will be chosen taking into account recent determinations of these parameters from the hyper-nucleus properties \cite{Fortin:2017cvt,Fortin:2017dsj,Fortin:2020qin}. In particular, we will study and compare  the properties of NS that only have nucleons, with the ones with $\Lambda$-hyperons, or with   $\Lambda$ and $\Xi^-$-hyperons, building sets of EOS with more than 15000 models which obey the two solar mass constraint. The recent NICER (Neutron star Interior Composition ExploreR) and LIGO/Virgo observations will be used to discuss the possible presence of hyperons.  These include:
the  measurement of the radius $12.71_{-1.19}^{+1.14}$ km and mass $1.34_{-0.16}^{+0.15}$ M$_\odot$  for the pulsar PSR J0030+0451 \cite{Riley:2019yda}, and the independent analysis which obtained the radius  $13.02_{-1.06}^{+1.24}$ km and the mass $1.44_{-0.14}^{+0.15}$ M$_\odot$ \cite{Miller:2019cac}; the recent measurement of the equatorial circumferential radius of the  pulsar PSR J0740+6620 with  mass $M=2.072_{-0.066}^{+0.067}$ M$_\odot$ and   $R=12.39^{+1.30}_{-0.98}$ km (68 $\%$ CI) \cite{Riley:2021pdl};  the radius estimation of $12.45 \pm 0.65$~km at 68\% CI for a $1.4 M_{\odot}$ NS from the simultaneous analysis of  NICER and  XMM-Newton X-ray observations; the tidal deformability determined from the detection of the the gravitational waves GW170817 \cite{LIGOScientific:2017ync,LIGOScientific:2018cki}.

We will first review the model used to perform the study, the choice of the parameters and the Bayesian inference approach. Results obtained in terms of corner plots for the model parameters, NMP, and  stellar matter properties  will be discussed. The 90\% CI for properties such as maximum mass configurations and respective radius, the radius and tidal deformability of NS with different masses,  the speed of sound or central baryonic densities will be calculated. Three scenarios will be compared: nuclear matter stellar matter, stellar matter with nucleons and $\Lambda$s and stellar matter with  nucleons and both  $\Lambda$s and $\Xi^-$s.  The two last scenarios are considered, only $\Lambda$s and $\Lambda$s and $\Xi^-$s, in order to quantify the effect of the onset hyperons, and, in particular, of a neutral hyperon and a second negatively charged hyperon. The onset of all the other possible hyperons gives rise to small amounts of new species and it is not expected that for densities as the ones occurring inside NS they will have a large impact. An EOS adequate to describe core collapse supernova matter or binary neutron star mergers which includes $\Lambda$s has been proposed in \cite{Banik:2014qja} and frequently used in simulations \cite{Radice:2017lry,Radice:2018pdn,Fujibayashi:2017puw}. However, it has also been shown that the inclusion of the other hyperonic species has non-negligible effects \cite{Marques:2017zju,Fortin:2017dsj}, and, therefore, we introduce both scenarios to discuss the possible differences. It will be shown that present observational constraints are compatible with the three scenarios.  The presence of hyperons implies larger radii and  smaller speeds of sound than expected for nucleonic matter. The measurement of a 1.4$M_\odot$ NS with a radius $\lesssim 12.5$ km or a  2 $M_\odot$ NS with a radius $\lesssim 11.5$ km would be incompatible with the presence of hyperons inside massive NS. 

{The paper is organized as follows, In Section \ref{model}, the field theoretical  DDH model for the  EOS at zero and finite temperatures is briefly reviewed, followed by a brief description of nuclear matter parameters (NMPs) in Section \ref{nmp} and Bayesian estimation of model parameters  in Section \ref{bayes}. The results of our calculation are discussed in Section \ref{sec:results}. Finally, the summary and conclusions are drawn in Section \ref{sec:con}.}

\section{Formalism}
\label{forma}
In this section,  we briefly review the relativistic mean-field description that will be used to generate {EOSs with different particle compositions: the nucleonic and the hyperonic sets.} 
We will also refer to  the statistical approach that will be used to estimate the model parameters.

\subsection{Model\label{model}}
The nuclear interaction will be described within a relativistic mean field (RMF) framework through the exchange of mesons, the isoscalar scalar $\sigma$-meson with mass $m_\sigma$, the isoscalar-vector $\omega$-meson with mass $m_\omega$ and the isovector-vector $\varrho$-meson with mass $m_\varrho$. {We will also include the vector $\phi$-meson with hidden strangeness and mass $m_\phi$ which will be responsible for the description of the hyperon-hyperon interaction.} Besides the nucleons, we will include the $\Lambda$-hyperon, the lightest neutral hyperon, and the $\Xi^-$, an hyperon with two strange quarks, which, therefore, sets in after the $\Lambda$-hyperon: it couples attractively to nuclear matter and  has a negative charge and, therefore,  replaces favorably the electrons reducing the stellar matter pressure. In order to simplify the treatment only these two hyperons are included: the $\Sigma$-meson seems to have a repulsive interaction in nuclear matter \cite{Gal:2016boi}, and the other isospin component of the $\Xi^-$ hyperon has a larger mass than the $\Lambda$. 
The model Lagrangian density has the form
\beq
\begin{aligned}
\mathcal{L}=& \sum_{B=n,p,\Lambda,\Xi^-}\bar{\Psi}_B\Big[\gamma^{\mu}\Big(i \partial_{\mu}-\Gamma_{B\omega} A_{\mu}^{(\omega)}-
\Gamma_{B\varrho} {\boldsymbol{{\bm t}_B}} \cdot \boldsymbol{A}_{\mu}^{(\varrho)} \\
&-\Gamma_{B\phi} A_{\mu}^{(\phi)}\Big) 
-\left(m_B-\Gamma_{B\sigma} \phi\right)\Big] \Psi_B
+ \frac{1}{2}\Big\{
\partial_{\mu} \phi \partial^{\mu} \phi \\
&-m_{\sigma}^{2} \phi^{2} \Big\} -\frac{1}{4} F_{\mu \nu}^{(\omega)} F^{(\omega) \mu \nu} 
+\frac{1}{2}m_{\omega}^{2} A_{\mu}^{(\omega)} A^{(\omega) \mu} \\
&-\frac{1}{4} F_{\mu \nu}^{(\phi)} F^{(\phi) \mu \nu} 
+\frac{1}{2}m_{\phi}^{2} A_{\mu}^{(\phi)} A^{(\phi) \mu} \\
&-\frac{1}{4} \boldsymbol{F}_{\mu \nu}^{(\varrho)} \cdot \boldsymbol{F}^{(\varrho) \mu \nu} 
+ \frac{1}{2} m_{\varrho}^{2} \boldsymbol{A}_{\mu}^{(\varrho)} \cdot \boldsymbol{A}^{(\varrho) \mu},
\end{aligned}
\label{lag}
\eeq
where  the baryons with bare mass $m_B$ are described by Dirac spinors $\Psi_B$,
$\gamma^\mu $ are the Dirac matrices and   $\bm{t}_B$ is the isospin operator for baryon $B$. $F^{(\omega, \phi)\mu \nu} = \partial^ \mu A^{(\omega, \phi)\nu} -\partial^ \nu A^{(\omega, \phi) \mu}$  and
$\boldsymbol{F}^{(\varrho)\mu \nu} = \partial^ \mu \boldsymbol{A}^{(\varrho)\nu} -\partial^ \nu \boldsymbol{A}^{(\varrho) \mu}
- \Gamma_{N\varrho} (\boldsymbol{A}^{(\varrho)\mu} \times \boldsymbol{A}^{(\varrho)\nu})$
are the vector meson field strength tensors,  $\Gamma_{B\sigma} $, $\Gamma_{B\omega}$, $\Gamma_{B\phi}$ and $\Gamma_{B\varrho}$ are the coupling constants of the baryons to the meson fields $\sigma$, $\omega$, $\phi$ and $\varrho$, respectively. For the density dependence of the couplings, we consider the functions  introduced in  \cite{Malik:2022zol}
\begin{equation}
  \Gamma_{B,M}(\rho) =\Gamma_{B,M,0} ~ h_M(x)~,\quad x = \rho/\rho_0~.
  \label{coupl}
\end{equation}
In this equation $\rho$ is the baryonic density, $\Gamma_{B,M,0}=x_{B M}\Gamma_{M,0}$ is the coupling  at saturation density $\rho_0$ of the meson $M$, $M \in \{ \sigma, \omega, \varrho, \phi \}$,  with the baryon $B\in \{n,\, p,\, \Lambda,\Xi^- \}$. For  the nucleons, i.e. $B=p,\, n$, the ratio $x_{B M}=1$ if $M\ne \phi$, and $x_{B \phi}=0$. {The other couplings will be discussed in Sec. \ref{sec:results}}.
In Eq. (\ref{coupl}) the function $h_M$ is given by
\begin{equation}
h_M(x) = \exp[-(x^{a_M}-1)],
\label{hm1}
\end{equation}
for the isoscalar mesons 
and  
\begin{equation}
h_\varrho(x) = \exp[-a_\varrho (x-1)] ~,
\label{hm2}
\end{equation}
for the $\varrho$ coupling, see \cite{Typel1999}.
We will describe static uniform matter in its ground state within the mean field approximation. The mesonic fields are replaced by their expectation values and only the time-like components of the vector fields, $\omega_0$, $\phi_0$ and the third isospin component of the $\varrho$ field $\varrho_3^0$,  survive. In the mean field approximation, the  Euler-Lagrange equations of all the fields are given by
\begin{eqnarray}
m_{\sigma}^{2} \sigma&=& \sum_{B=n,p,\Lambda,\Xi^-}\Gamma_{B\sigma} \bar{\psi}_B \psi_B, \\
\quad m_{\omega}^{2} \omega_{0}&=& \sum_{B=n,p,\Lambda,\Xi^-}\Gamma_{B\omega} \bar{\psi}_B \gamma_{0} \psi_B, \\
\quad m_{\varrho}^{2} \varrho_{3}^0&=&\frac{1}{2} \sum_{B=n,p,\Xi^-}\Gamma_{B\varrho} \bar{\psi}_B \gamma_{0} \tau_{3} \psi_B
\end{eqnarray}
The nucleon number density $\rho=<\bar{\psi} \gamma_{0} \psi>$ and scalar density $\rho_s=<\bar{\psi} \psi>$ 
at zero temperature are defined as,
\begin{eqnarray}
\rho&=&\frac{\gamma}{2\pi^{2}} \sum_{B=p,n,\Lambda,\Xi^-}\int_{0}^{k_{F_B}} 
k^2\,d k, \\
\rho_{s}&=&\frac{\gamma}{2 \pi^{2}} \sum_{B=p,n,\Lambda,\Xi^-}\int_{0}^{k_{F_B}} 
\frac{m_B^{*} k^2}{\sqrt{m_B^{* 2}+k^{2}}} \,dk,
\end{eqnarray}
where $\gamma$ is the spin degeneracy factor, $k_{F_B}$  is  the  Fermi momentum of baryon $B$ with effective nucleon mass is $m_B^{*}=m_B - \Gamma_{B\sigma} \sigma$ and  chemical potential  $\mu_B=\nu_{B}+ \Gamma_{B\omega} \omega_{0}+ \Gamma_{B\varrho} \tau_{3 B} \varrho_{3}^0 +\Sigma^{r} $, with  $\tau_{3 B}$  the isospin projection. The rearrangement term $\Sigma^{r}$  assures thermodynamic consistency. It arises due to the density-dependence of the couplings and is expressed as
\begin{eqnarray}
\Sigma^{r}&=\sum_{B=n,p,\Lambda,\Xi^-}\Bigg[-\frac{\partial \Gamma_{B\sigma}}{\partial \rho_{B}} \sigma \rho_{sB}+\frac{\partial \Gamma_{B\omega}}{\partial \rho_{B}} \omega_{0} \rho_{B}
\nonumber \\ &+\frac{\partial \Gamma_{B\phi}}{\partial \rho_{B}} \phi_{0} \rho_{B} 
+\frac{\partial \Gamma_{B\varrho}}{\partial \rho_{B}} \tau_{3 B} \rho_{3}^0 \rho_{B}\Bigg].
\end{eqnarray}
The energy density is defined as,
\begin{eqnarray}
\varepsilon&=&\frac{1}{\pi^{2}} \sum_{B={n},{p},\Lambda,\Xi^-} \int_{0}^{k_{F_B}} k^{2} \sqrt{k^{2}+m_B^{* 2}} d k + \frac{1}{2} m_{\sigma}^{2} \sigma^{2} \nonumber \\
&+&\frac{1}{2} m_{\omega}^{2} \omega_{0}^{2}+\frac{1}{2} m_{\phi}^{2} \phi_{0}^{2}+\frac{1}{2} m_{\varrho}^{2} (\varrho_{3}^0)^{2}+ {\varepsilon_{lep}}.
\end{eqnarray}
In this expression the last term takes into account  the leptonic  contribution of both electrons and muons.
The pressure $P$ is calculated from  the Euler relation,
\begin{equation}
P={\sum_{i=n,p,\Lambda,\Xi^-,e,\mu}} \mu_{i} \rho_{i}-\varepsilon,
\end{equation}
{where $\mu_i$ and $\rho_i$ are, respectively, the chemical potential and the number density of particle $i$.}

We consider that cold $\beta$-equilibrium matter is an electrically neutral matter. These two conditions impose some relation between the particle chemical potentials and densities, in particular
\bea 
\label{ch02}
\mu_B=\mu_n- q_B\mu_e \quad {\rm and}\qquad\mu_e=\mu_\mu,
\eea 
with $q_B$ the electric charge of baryon $B$ and $\mu_n$,  $\mu_e$, $\mu_\mu$  the  neutron, electron and muon chemical potential, respectively; and
\bea 
\label{ch01}
\rho_p=\rho_{\Xi^-}+\rho_e+\rho_{\mu}.
\eea 
with $\rho_p$, $\rho_{\Xi^-}$, $\rho_e$ and $\rho_\mu$ the number density of  protons, cascades, electrons and muons, respectively.

For the crust EOS, the procedure described in \cite{Malik:2022zol} will be considered:  the outer crust is described by the Bethe-Pethick-Sutherland (BPS) EOS;  
The outer crust and the core are joined using a polytropic function 
 \cite{Carriere:2002bx} $p(\varepsilon)=a_1 + a_2 \varepsilon^{\gamma}$, where the parameters $a_1$ and $a_2$ are determined in such a way that 
the EOS for the inner crust  joins the upper layer of the outer crust at ($\rho=10^{-4}$ fm$^{-3}$) to  the core EOS at ($\rho=0.04$ fm$^{-3}$). The polytropic index $\gamma$ is taken to be equal to $4/3$. It has been discussed in \cite{Fortin2016,Pais:2016xiu} that this approximation introduces an uncertainty on the radius of the low mass NS. In order to decrease the uncertainty, the core EOS is used also to described the upper layers of the inner crust, since  the inner crust EOS does not differ much from the homogeneous EOS for densities close to the transition to the core  \cite{Avancini2008}. In \cite{Malik:2022zol} we have estimated that this procedure introduces for models with a symmetry energy compatible with the $\chi$EFT PNM EOS an uncertainty  not greater than $100-200$m.

\subsection{Nuclear matter parameters (NMPs) \label{nmp}}
{The EOS of nuclear matter can be decomposed into two parts 
with a good approximation:} (i) the EOS for symmetric nuclear matter $\epsilon(\rho,0)$ (ii) a term involving the symmetry energy coefficient $S(\rho)$ and the asymmetry $\delta$,
\bea
 \epsilon(\rho,\delta)\simeq \epsilon(\rho,0)+S(\rho)\delta^2,
 \label{eq:eden}
\eea 
{where  $\epsilon$ is the energy per nucleon at a given density  $\rho$  and isospin asymmetry} $\delta=(\rho_n-\rho_p)/\rho$.
{The EOS can be transformed} in terms of various bulk nuclear matter properties {of order $n$ at saturation density: (i) for the symmetric nuclear matter, the energy per nucleon $\epsilon_0=\epsilon(\rho_0,0)$ ($n=0$), the incompressibility coefficient $K_0$ ($n=2$), the
skewness  $Q_0$ ($n=3$),  and  the kurtosis $Z_0$ ($n=4$), respectively, given by
\begin{equation}
X_0^{(n)}=3^n \rho_0^n \left (\frac{\partial^n \epsilon(\rho, 0)}{\partial \rho^n}\right)_{\rho_0}, \, n=2,3,4;
\label{x0}
\end{equation}
(ii) for the symmetry energy,  the symmetry energy at saturation 
 $J_{\rm sym,0}$ ($n=0$), 
\begin{equation}
J_{\rm sym,0}= S(\rho_0);~ ~ S(\rho)=\frac{1}{2} \left (\frac{\partial^2 \epsilon(\rho,\delta)}{\partial\delta^2}\right)_{\delta=0},
\end{equation}
the slope $L_{\rm sym,0}$ ($n=1$),  the curvature $K_{\rm sym,0}$ ($n=2$),  the skewness $Q_{\rm sym,0}$ ($n=3$), and  the kurtosis $Z_{\rm sym,0}$ ($n=4$), 
respectively, defined as
\begin{equation}
X_{\rm sym,0}^{(n)}=3^n \rho_0^n \left (\frac{\partial^n S(\rho)}{\partial \rho^n}\right )_{\rho_0},\, n=1,2,3,4.
\label{xsym}
\end{equation}
}

\subsection{Bayesian estimation of Model Parameters\label{bayes}}
A Bayesian inference approach  estimates the parameter values by updating a prior belief of the model parameters (i.e., prior distribution) with new evidence (i.e., observed/fit data) via optimizing a likelihood function, resulting in a posterior distribution. The marginalized posterior distributions of model parameters enables one to carry out a detailed statistical analysis of all the predicted quantities by the models. The posterior distributions of the model parameters $\theta$ in Bayes’ theorem can be written as,
\begin{equation}
P(\bm{\theta} |D ) =\frac{{\mathcal L } (D|\bm{\theta}) P(\bm {\theta })}{\mathcal Z},\label{eq:bt}
\end{equation}
where $\bm{\theta}$  and $D$ denote the set of model parameters and the fit data.  $P(\bm {\theta })$ in Eq. (\ref{eq:bt}) is the prior for the  model parameters and $\mathcal Z$ is the evidence. The type of prior can be chosen with the preliminary knowledge of the model parameters. One can choose it to be a uniform prior, which has been used as a baseline for many analyses. The $P(\bm{\theta} |D )$ is the joint posterior distribution of the parameters, $\mathcal L (D|\bm{\theta})$ is the likelihood function.
The posterior distribution  of a given parameter can be obtained by marginalizing $P(\bm{\theta} |D )$ over  the remaining parameters. The marginalized posterior distribution for a  parameter $\theta_i$
is obtained as,
\begin{equation}
 P (\theta_i |D) = \int P(\bm {\theta} |D) \prod_{k\not= i }d\theta_k. \label{eq:mpd}
\end{equation}
We use a Gaussian likelihood function defined as, 
\bea
{\mathcal L} (D|\bm{\theta})&=&\prod_{j} 
\frac{1}{\sqrt{2\pi\sigma_{j}^2}}e^{-\frac{1}{2}\left(\frac{d_{j}-m_{j}(\bm{\theta)}}{\sigma_{j}}\right)^2}. 
\label{eq:likelihood}  
\eea
Here the index $ j$ runs over all the data, $d_j$ and $m_j$ are the data and corresponding model values, respectively.  The $\sigma_j$ are the adopted uncertainties.

{{\it Sampling-} We employ the Nested Sampling technique to sample the prior distribution for this work. It was first proposed in Ref. \cite{Skilling2004} and suitable for low dimensional problem. In Nested Sampling,  the posterior is broken into many nested “slices” with starting {\it"n-live"} points,  samples are generated from each of them and then recombined  to reconstruct the original distribution. We invoke the {\it Pymultinest} sampler \cite{Buchner:2014nha,buchner2021nested} in the python library and generate samples for starting 4000 {\it"n-live"} points.}

\begin{table}[]
\caption{{The uniform prior (P) setup considered for the DDH model parameters in this work. The parameters ’min’ and ’max’ denote the minimum and maximum values for the uniform distribution.}}
\label{tab2}
\setlength{\tabcolsep}{15.5pt}
\renewcommand{\arraystretch}{1.1}
\begin{tabular}{cccc}
\toprule
\multirow{2}{*}{No} & \multirow{2}{*}{Parameters}  & \multicolumn{2}{c}{P} \\ \cline{3-4} 
                    &                                & min  & max  \\ \hline
1                   & $\Gamma_{\sigma,0}$                              & 6.5  & 13.5   \\
2                   & $\Gamma_{\omega,0}$                              & 7.5  & 14.5   \\
3                   & $\Gamma_{\varrho,0}$                        & 2.5    & 8.0  \\
4                   & $a_{\sigma}$                               & 0.0    & 0.30 \\
5                   & $a_{\omega}$                             & 0.0    & 0.30 \\
6                   & $a_{\varrho}$                            & 0.0 & 1.30    \\ 
7                   & $x_{\sigma\Lambda}$                      & 0.609 & 0.622  \\
8                   & $x_{\sigma\Xi^-}$                        & 0.309 & 0.322  \\ \hline
\hline
\end{tabular}
\end{table}

\section{Results\label{sec:results}}
{We consider that it is important to introduce two hyperons, a neutral one and a negatively charged one: the neutral one will mitigate the  neutron pressure as soon as its chemical potential is equal to the neutron chemical potential; the negatively charged one is  important because its onset is favored to relieve the electron  pressure.  Other hyperon species could set in, but  it is expected that their  overall influence in the NS properties would be small and, in order to keep the picture the simplest possible,} {we restrict the present study to the inclusion of the $\Lambda$ and $\Xi^-$ hyperons as exotic degrees of freedom in NS matter.}
The reason to choose these two hyperons are:  $\Lambda$ is the first hyperon to set in, if we consider that the $\Sigma$-hyperon couples repulsively to nuclear matter,  as the non-existence of $\Sigma$-hypernuclei seems to indicate \cite{Gal:2016boi};  the second hyperon is not so clear, but considering for the $\Sigma$-hyperon a repulsive optical potential, its onset is frequently shifted to densities above the onset of the $\Xi^-$ \cite{Weissenborn:2011kb,Fortin2016,Fortin:2020qin,Stone:2019blq}. 

In our study, we adopt the SU(6) values for the couplings to vector-isoscalar mesons:
\bea
	&&g_{\omega\Xi^-}=\frac{1}{3} g_{\omega N} = \frac{1}{2} g_{\omega\Sigma} \ ,
		\label{eq:SU6-relation2}
\eea
\bea
         && g_{\phi\Xi^-} = 2 g_{\phi\Sigma} =- \frac{2\sqrt{2}}{3} g_{\omega N} 
	\label{eq:SU6-relation2b}
\eea
and assume 
\bea	
g_{\rho B} = g_{\rho N}  
	\label{eq:SU6-relation3}
\eea
for the $\rho$-meson, the isospin properties of the different baryons being taken into account in the Lagrangian density (\ref{lag}). With this choice of couplings to the vector mesons, 
 the $\sigma$-meson coupling to the $\Lambda$-hyperon was fitted to hypernuclei properties for several RMF models in \cite{Fortin:2017cvt,Fortin:2017dsj,Providencia:2018ywl}. Expressing this coupling parameter in terms of the coupling to the nucleon, $g_{\sigma\Lambda}=x_{\sigma\Lambda}\, g_{\sigma N}$, the fraction $x_{\sigma\Lambda}$ was determined for each model and values between 0.609 and 0.622 have been obtained. The calibration of the $g_{\sigma\Xi^-}$ coupling was performed in
  \cite{Fortin:2020qin} with the same assumption for the vector mesons. The coupling of the $\sigma$-meson to the $\Xi^-$ was fitted to the experimental binding energy of two single-$\Xi$ hypernuclei, $^{15}_{\Xi}$C and $^{12}_{\Xi}$Be, and values between 0.309 and 0.321 have been obtained for $x_{\sigma\Xi^-}$.
In our study, we allow the fractions $x_{\sigma\Lambda}$ and $x_{\sigma\Xi^-}$ to vary in the intervals [0.609,0.622] and [0.309,0.322], respectively. 

We generate the distribution of DDH model parameters for each set within a Bayesian parameter estimation approach considering a given set of fit data related with the nuclear saturation properties, the pure neutron matter EOS calculated from a precise N$^3$LO calculation in $\chi$EFT and the lowest bound of NS observational maximum mass. The three different sets of EOSs considered for the present study are: (i) a set of only nucleonic EOS (DDB) \cite{Malik:2022zol}, { prior defined by numbers 1 to 6 of Table \ref{tab2}}, (ii) a set of EOS that includes nucleons and the $\Lambda$ hyperon (DDB$\Lambda$), { prior defined by numbers 1 to 7 of Table \ref{tab2}}, and (iii) a set of EOS with nucleons and  the $\Lambda$ and $\Xi^-$ hyperons (DDB$\Lambda\Xi^-$),{ prior defined by numbers 1 to 8 of Table \ref{tab2}} . {In particular, the last two sets are generated in this work and the number of EOS in each set is $\sim 15,000$. The marginalized posterior distributions of  model parameters in each EOS set enable us to  perform a detailed statistical analysis of the nuclear matter parameters and the NS properties. The marginalized posterior distributions of the model parameters,} applying a  Bayesian estimation, 
requires the definition of  the likelihood, of the fit data and of the priors for the model parameters. The likelihood has been defined in Sec. \ref{bayes}, see Eq. (\ref{eq:likelihood}). The list of fit data considered for the present study is presented in Table \ref{tab1}. 
\begin{table}[]
\centering
 \caption{The constraints used in the Bayesian inference  of the model parameters to generate  the nucleonic set DDH and  the two hyperonic EOS sets DDB$\Lambda$ and DDB$\Lambda\Xi^-$: 
 the  binding energy per nucleon $\epsilon_0$,  {the} incompressibility $K_0$ and   the symmetry energy  $J_{\rm sym,0}$ calculated  at the nuclear saturation density  $\rho_0$, including an 1$\sigma$ uncertainty;  the pressure of pure neutron matter  PNM determined at the densities 0.08, 0.12 and 0.16~fm$^{-3}$ from  a N$^3$LO calculation in $\chi$EFT \cite{Hebeler2013}, with {2 $\times$} N$^3$LO {uncertainty}  in the likelihood.}
  \label{tab1}
 \setlength{\tabcolsep}{5.5pt}
      \renewcommand{\arraystretch}{1.1}
\begin{tabular}{cccc}
\hline 
\hline 
\multicolumn{4}{c}{Constraints}                                                        \\
\multicolumn{2}{c}{Quantity}                     & Value/Band  & Ref     \\ \hline
\multirow{3}{*}{\shortstack{NMP \\  {[}MeV{]} }} 
& $\rho_0$ & $0.153\pm0.005$ & \cite{Typel1999}    \\
& $\epsilon_0$ & $-16.1\pm0.2$ & \cite{Dutra:2014qga}   \\
                               & $K_0$           & $230\pm40$   & \cite{Shlomo2006,Todd-Rutel2005}    \\
                              & $J_{\rm sym, 0}$           & $32.5\pm1.8$  & \cite{Essick:2021ezp}   \\
                              
                               &                 &                &                                                   \\
  \shortstack{PNM \\ {[}MeV fm$^{-3}${]}}                  & $P(\rho)$       & $2\times$ N$^{3}$LO    & \cite{Hebeler2013}   \\
\shortstack{NS mass \\ {[}$M_\odot${]}}        & $M_{\rm max}$   & $>2.0$     &  \cite{Fonseca:2021wxt}      \\ 
\hline 
\end{tabular}
\end{table}

In Fig. \ref{T:fig1} we plot the distribution of model parameters for all the three different cases: (i) DDB, (ii) DDB$\Lambda$ and, (iii) DDB$\Lambda\Xi^-$. The confidence ellipses  for two dimensional posterior distributions are plotted with 1$\sigma$, 2$\sigma$ and 3$\sigma$ confidence intervals (CI). The marginalized 1D distribution of each and every individual parameter is shown in the diagonal blocks of the corner plot. The vertical lines shown in the 1D distributions indicate the 68\% min, median and 68\% max CI of the model parameters, respectively. The inclusion of hyperons tends to make the EOS softer.  Imposing simultaneously the 2 M$_\odot$ NS maximum mass constraint, which is considered in every case, pushes the EOS to a stiffer regime. It can be noted from the figure that {only the  values for the couplings $\Gamma_{\sigma,0}$ and $\Gamma_{\omega,0}$  closer to the upper limit of the prior, and the values of $a_\sigma$ and $a_\omega$  parameters closer to the lower limit of the prior (see Table \ref{tab2})}  can sustain the 2 M$_\odot$ constraint,  when hyperons are included.  As a consequence, the different CI for the parameters $\Gamma_{\sigma,0}$, $\Gamma_{\omega,0}$, $a_\sigma$ and $a_\omega$ get shrinked with respect to DDB in the case of DDB$\Lambda$ and further shrinked in the case of DDB$\Lambda\Xi^-$. The $\Xi^-$ makes the EOS  in DDB$\Lambda\Xi^-$ set  even softer than EOS belonging to the DDB$\Lambda$ set. 
However, it should be referred that the parameters defining the $\rho$ meson couplings are not effected by the presence of hyperons. As mentioned earlier,  the priors of the parameters $x_{\sigma\Lambda}$ and $x_{\sigma\Xi^-}$
are taken to be uniform distribution in the intervals [0.609,0.622] and [0.309,0.322], respectively. The posterior obtained for those two parameters are also uniform and are not shown in the figure. 
\begin{figure}
\includegraphics[width=0.45\textwidth]{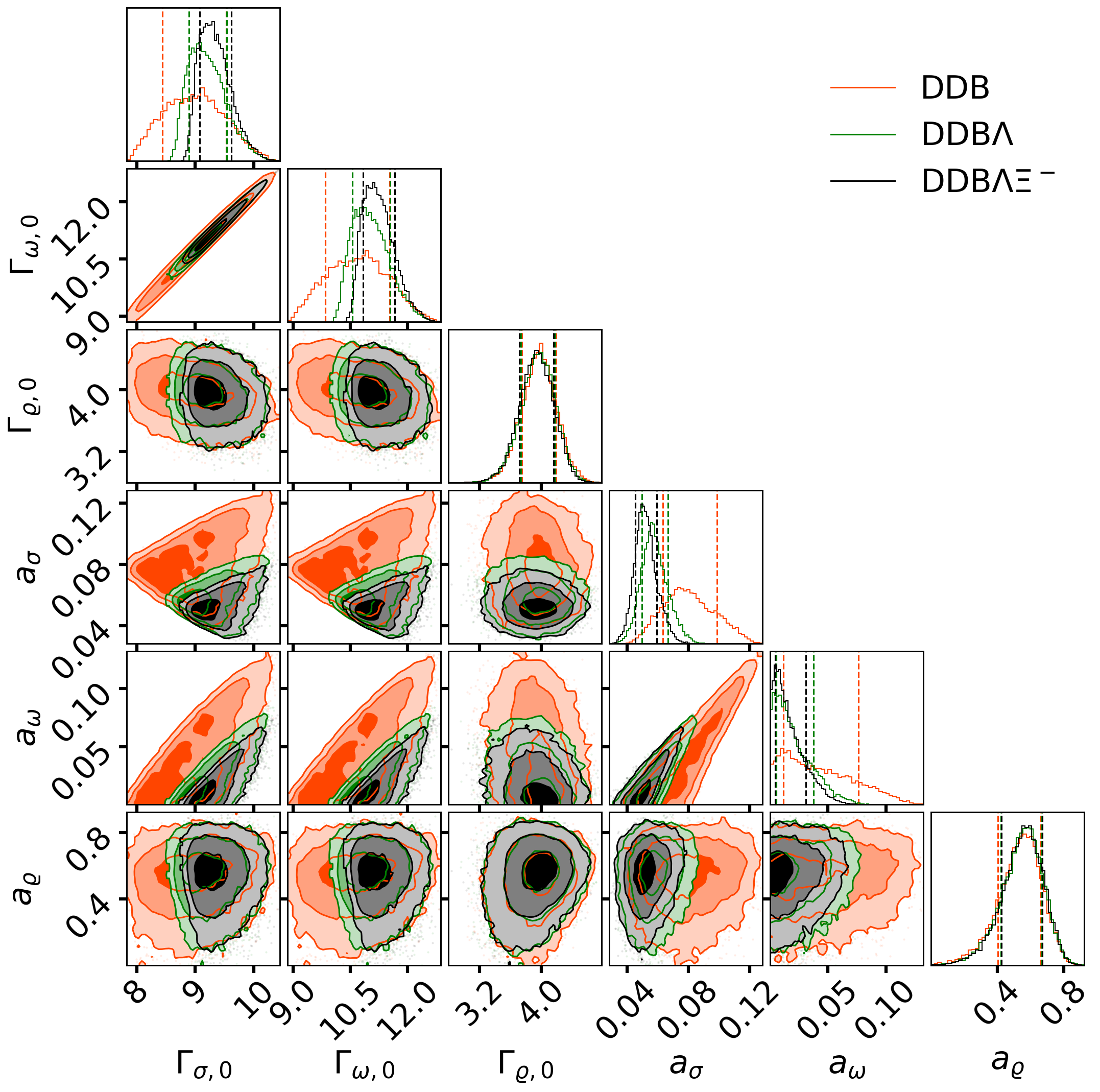}
\caption{
Corner plot for the marginalized posterior distributions of the model parameters. The 68\% CI are indicated with vertical lines for the  DDB (red), DDB with $\Lambda$ hyperon (green) and DDB with $\Lambda$ and $\Xi^-$ hyperon (black) sets.  Also shown are the  1$\sigma$, 2$\sigma$ and 3$\sigma$ CI (tonalities from dark to light)  of the two dimensional posterior distributions  ellipses.
\label{T:fig1}} 
\end{figure}

{We now proceed to analyze the ensemble of EOSs obtained for the three different sets DDB, DDB$\Lambda$ and DDB$\Lambda\Xi^-$. In Fig. \ref{T:particle}, (left) we plot the particle fraction along with associated 90\% CI for proton, $\Lambda$ and $\Xi^-$ hyperon, (right) the pressure for the three different sets as a function of baryon density $\rho$. In the right panel, we also plot the constraints for NS matter EOS obtained from GW170817 analysis \cite{LIGOScientific:2018cki} for comparison.} {The onset of hyperons has a strong effect on the proton fraction: if only $\Lambda$-hyperons are considered the proton fraction decreases to fractions well below 10\% while the $\Lambda$ fraction can reach fractions as high as $\sim50$\%; the inclusion of $\Xi^-$ increases slightly the proton fraction and reduces the $\Lambda$ fraction to values below 30\%. A decrease of the proton fraction disfavors the nucleonic Urca process. However, the onset of hyperons opens other Urca channels \cite{Fortin2016,Fortin:2020qin}. The 90\% CI of  $\beta$-equilibrium  pressure as a function of baryon density obtained for all the three, DDB, DDB$\Lambda$ and DDB$\Lambda\Xi^-$ are fully compatible with the GW170817 constraints. It is to be noted that the analysis performed for GW170817 did not impose the 2$M_\odot$ constraint.}
\begin{figure*}
\includegraphics[width=0.48\textwidth]{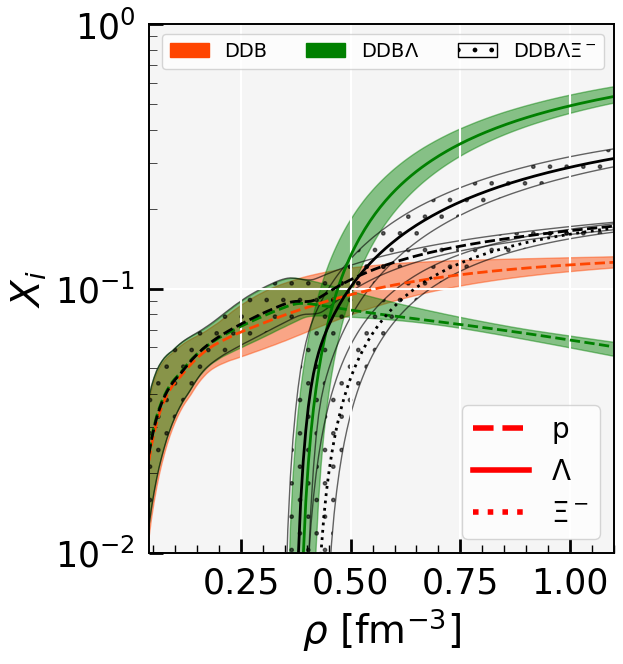}
\includegraphics[width=0.48\textwidth]{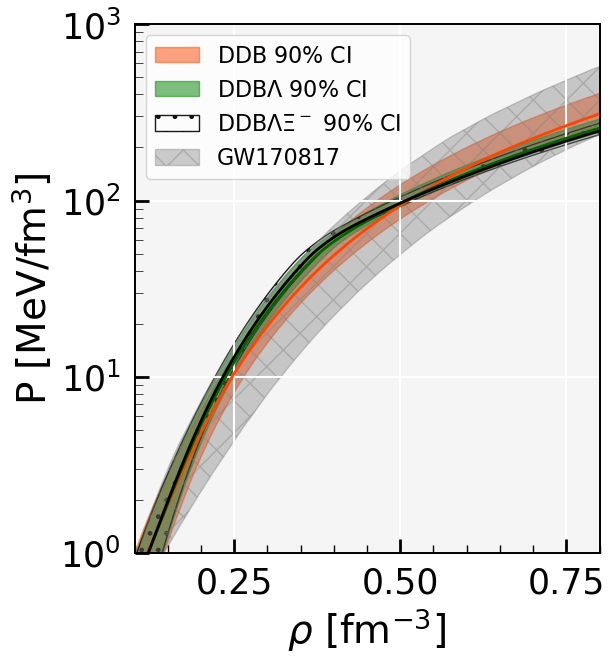}
\caption{{The median and  90\% CI  for the particle fractions $X_i$  (protons (p) "dashed line", lambdas ($\Lambda$) "solid line" and cascades ($\Xi^-$) "dotted line") (left panel) and for the NS matter pressure  (right panel) versus baryonic density $\rho$ for the DDB (red), DDB$\Lambda$ (green) and DDB$\Lambda\Xi^-$ (dotted) sets. 
The  GW170817 constraints for the pressure  \cite{LIGOScientific:2018cki} are also shown.}} 
\label{T:particle}
\end{figure*}


\begin{table*}[]
\caption{
The median values and the associated  90\% CI of the NMPs defined in Eq. (\ref{x0}) and (\ref{xsym}), and NS properties, the gravitational mass $M_{\rm max}$, baryonic mass  $M_{\rm B, max}$, radius $R_{\rm max}$, central energy density $\varepsilon_c$, central number density for baryon $\rho_c$ and square of central speed-of-sound $c_s^2$ of the maximum mass NS, the radius   $R_{{\rm M}_i}$ and  the dimensionless tidal deformability $\bar \Lambda_{{\rm M}_i}$ for NS mass ${\rm M}_i \in [1.4,1.6,1.8,2.08]$ $M_\odot$, and the effective tidal deformability $\tilde \Lambda$ for the GW170817 merger with $q=1$ ($q$ is the mass ratio of NSs involved in binary merger) {obtained for the DDB, DDB$\Lambda$ and DDB$\Lambda\Xi^-$ sets. The $X_\Lambda^c$ and $X_{\Xi^-}^c$ are the fraction of $\Lambda$ and $\Xi^-$ at the core of the maximum mass NS, respectively.}  
\label{tab4}}
 \setlength{\tabcolsep}{5.2pt}
      \renewcommand{\arraystretch}{1.4}
\centering
\begin{tabular}{cccccccccccccc}
\hline
\hline
\multicolumn{2}{c}{\multirow{3}{*}{Quantity}}    & \multirow{3}{*}{Units} & \multicolumn{3}{c}{DDB}                                 &  & \multicolumn{3}{c}{DDB$\Lambda$}                        &  & \multicolumn{3}{c}{DDB$\Lambda\Xi^-$}                     \\ \cline{4-6} \cline{8-10} \cline{12-14} 
\multicolumn{2}{c}{}                             &                        & \multirow{2}{*}{median} & \multicolumn{2}{c}{$90\%$ CI} &  & \multirow{2}{*}{median} & \multicolumn{2}{c}{$90\%$ CI} &  & \multirow{2}{*}{median} & \multicolumn{2}{c}{$90\%$ CI} \\ \cline{5-6} \cline{9-10} \cline{13-14} 
\multicolumn{2}{c}{}                             &                        &                         & min           & max           &  &                         & min           & max           &  &                         & min           & max           \\ \hline
\multirow{10}{*}{NMP} & $\rho_0$                 & fm$^{-3}$              & $0.153$                 & $0.147$       & $0.158$       &  & $0.152$                 & $0.147$       & $0.157$       &  & $0.152$                 & $0.147$       & $0.157$       \\
                      & $\varepsilon_0$          & \multirow{9}{*}{MeV}   & $-16.10$                & $-16.41$      & $-15.80$      &  & $-16.09$                & $-16.40$      & $-15.79$      &  & $-16.09$                & $-16.39$      & $-15.79$      \\
                      & $K_0$                    &                        & $231$                   & $201$         & $276$         &  & $260$                   & $235$         & $300$         &  & $273$                   & $249$         & $309$         \\
                      & $Q_0$                    &                        & $-109$                  & $-256$        & $130$         &  & $58$                    & $-82$         & $285$         &  & $138$                   & $-4$          & $352$         \\
                      & $Z_0$                    &                        & $1621$                  & $735$         & $2020$        &  & $1491$                  & $826$         & $1755$        &  & $1420$                  & $809$         & $1668$        \\
                      & $J_{\rm sym,0}$          &                        & $32.19$                 & $29.38$       & $34.81$       &  & $32.22$                 & $29.50$       & $34.79$       &  & $32.15$                 & $29.48$       & $34.83$       \\
                      & $L_{\rm sym,0}$          &                        & $41.26$                 & $22.79$       & $65.16$       &  & $42.37$                 & $24.84$       & $65.18$       &  & $43.52$                 & $25.92$       & $65.27$       \\
                      & $K_{\rm sym,0}$          &                        & $-116$                  & $-150$        & $-73$         &  & $-104$                  & $-134$        & $-64$         &  & $-97$                   & $-127$        & $-59$         \\
                      & $Q_{\rm sym,0}$          &                        & $966$                   & $317$         & $1469$        &  & $981$                   & $355$         & $1461$        &  & $966$                   & $356$         & $1447$        \\
                      & $Z_{\rm sym,0}$          &                        & $-6014$                 & $-11564$      & $-1911$       &  & $-6548$                 & $-11947$      & $-2519$       &  & $-6740$                 & $-12109$      & $-2768$       \\
                      &                          &                        &                         &               &               &  &                         &               &               &  &                         &               &               \\
\multirow{15}{*}{NS}  & $M_{\rm max}$            & M $_\odot$             & $2.144$                 & $2.021$       & $2.355$       &  & $2.061$                 & $2.012$       & $2.167$       &  & $2.049$                 & $2.010$       & $2.137$       \\
                      & $M_{\rm B, max}$         & M $_\odot$             & $2.552$                 & $2.386$       & $2.835$       &  & $2.414$                 & $2.350$       & $2.552$       &  & $2.391$                 & $2.340$       & $2.504$       \\
                      & $c_{s}^2$                & $c^2$                  & $0.65$                  & $0.53$        & $0.72$        &  & $0.49$                  & $0.45$        & $0.52$        &  & $0.48$                  & $0.44$        & $0.50$        \\
                      & $\rho_c$                 & fm$^{-3}$              & $0.946$                 & $0.879$       & $0.970$       &  & $0.951$                 & $0.880$       & $0.964$       &  & $0.928$                 & $0.870$       & $0.961$       \\
                      & $\varepsilon_{c}$        & MeV fm$^{-3}$          & $1173$                  & $1130$        & $1173$        &  & $1173$                  & $1086$        & $1173$        &  & $1130$                  & $1065$        & $1173$        \\
                      & $X_\Lambda^c$        &           & $-$                  & $-$        & $-$        &  & $0.466$                  & $0.442$        & $0.487$        &  & $0.269$                  & $0.251$        & $0.287$        
                      \\
                      & $X_{\Xi^-}^c$        &           & $-$                  & $-$        & $-$        &  & $-$                  & $-$        & $-$        &  & $0.153$                  & $0.146$        & $0.159$        
                      \\
                      & $R_{\rm max}$            & \multirow{5}{*}{km}    & $11.09$                 & $10.56$       & $11.74$       &  & $11.58$                 & $11.30$       & $11.99$       &  & $11.75$                 & $11.52$       & $12.09$       \\
                      & $R_{1.4}$                &                        & $12.62$                 & $12.07$       & $13.21$       &  & $12.98$                 & $12.62$       & $13.43$       &  & $13.12$                 & $12.79$       & $13.52$       \\
                      & $R_{1.6}$                &                        & $12.53$                 & $11.95$       & $13.18$       &  & $12.97$                 & $12.61$       & $13.41$       &  & $13.13$                 & $12.82$       & $13.52$       \\
                      & $R_{1.8}$                &                        & $12.36$                 & $11.72$       & $13.11$       &  & $12.83$                 & $12.46$       & $13.32$       &  & $13.02$                 & $12.71$       & $13.45$       \\
                      & $R_{2.075}$              &                        & $12.01$                 & $11.10$       & $12.93$       &  & $12.35$                 & $11.80$       & $13.04$       &  & $12.52$                 & $12.00$       & $13.12$       \\
                        & $R_{2.2}$       &      & $12.15$ & $11.43$ & $12.86$ & & $12.60$ & $12.18$ & $13.09$ & & $12.66$ & $12.45$ & $12.95$      \\
                      & $\bar \Lambda_{1.4}$          & \multirow{5}{*}{-}     & $454$                   & $339$         & $625$         &  & $568$                   & $480$         & $708$         &  & $615$                   & $534$         & $744$         \\
                      & $\bar \Lambda_{1.6}$          &                        & $185$                   & $132$         & $269$         &  & $243$                   & $202$         & $308$         &  & $268$                   & $231$         & $328$         \\
                      & $\bar \Lambda_{1.8}$          &                        & $79$                    & $52$          & $125$         &  & $106$                   & $85$          & $142$         &  & $120$                   & $101$         & $153$         \\
                      & $\bar \Lambda_{2.075}$        &                        & $22$                    & $11$          & $43$          &  & $27$                    & $18$          & $45$          &  & $30$                    & $20$          & $47$          \\
                      & $\tilde \Lambda_{q=1.0}$ &                        & $454$                   & $339$         & $625$         &  & $568$                   & $480$         & $708$         &  & $615$                   & $534$         & $744$         \\ \hline
\end{tabular}
\end{table*}
 We perform next a statistical analysis of the NMPs and neutron star properties, namely, mass, radius, central speed-of-sound square and energy density,  and dimensionless tidal deformability  using the calculated marginalized posterior distributions of the model parameters for three sets (DDB, DDB$\Lambda$ and DDB$\Lambda\Xi^-$). The NS masses and radii were calculated from the TOV equations \cite{TOV1,TOV2}  and the dimensionless tidal deformability ($\bar \Lambda = \frac{2}{3}~ k_2 (M/R)^{-5}$, $k_2$ is the second Love number) from the equations obtained in \cite{Hinderer2008}. In Table \ref{tab4}, we present the median values and the associated 90\% CI uncertainties of the NMPs and of some NS properties, namely, the gravitational mass M$_{\rm max}$, the baryonic mass $M_{\rm B, max}$, the central speed-of-sound square $c_s^2$, the central energy density $\varepsilon_{c}$, {fraction of $\Lambda$ ($X_\Lambda^c$) and $\Xi^-$ ($X_{\Xi^-}^c$) hyperon at the center of the NS,} and  the radius $R_{\rm max}$ of maximum mass NS, as well as the radius   and  the dimensionless tidal deformability for {1.4, 1.6, 1.8 and 2.075 $M_\odot$ NS  (also the radius of 2.2 $M_\odot$ NS and combined tidal deformability $\tilde\Lambda_{q=1}$ for mass fraction $q=1$ in a binary NS meager) obtained for DDB, DDB$\Lambda$ and DDB$\Lambda\Xi^-$. In Figs. \ref{T:fig3a} and \ref{T:fig3b} are given the corner plots for the  same {few} quantities, respectively,  NMPs  and NS properties.} 
\begin{figure*}
\includegraphics[width=0.95\textwidth]{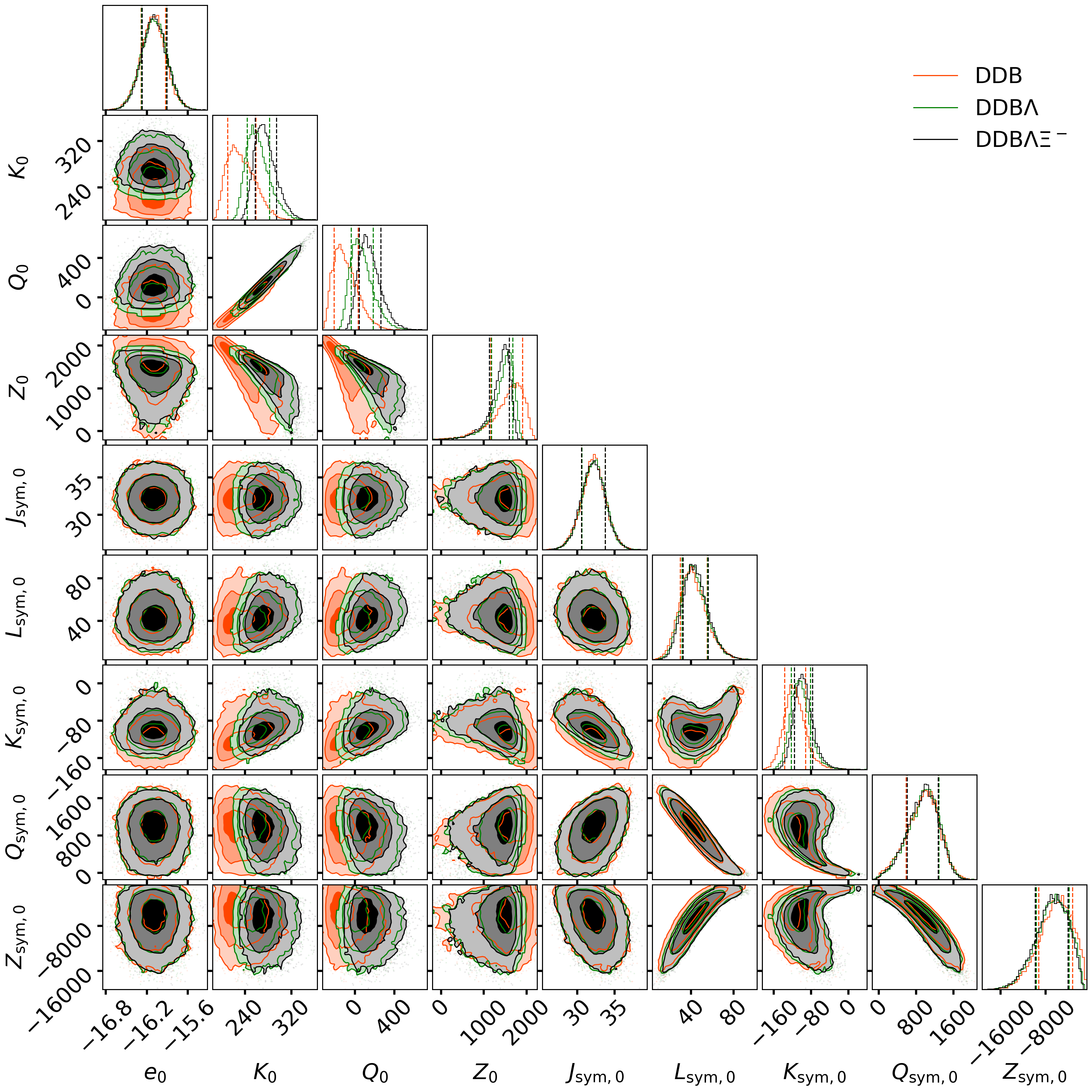}
\caption{Corner plot for the NMPs marginalized posterior distributions (MeV) calculated for the  DDB (red), DDB with $\Lambda$ hyperon (green) and DDB with $\Lambda$ plus $\Xi^-$ hyperon (black) sets, see Eqs. (\ref{x0}) and (\ref{xsym}) \label{T:fig3}. The vertical lines indicate the 68\% CI  and the different tonalities from dark to light indicate, respectively, the 1$\sigma$, 2$\sigma$, and 3$\sigma$ CI.
\label{T:fig3a}
}
\end{figure*}
\begin{figure*}
\includegraphics[width=0.95\textwidth]{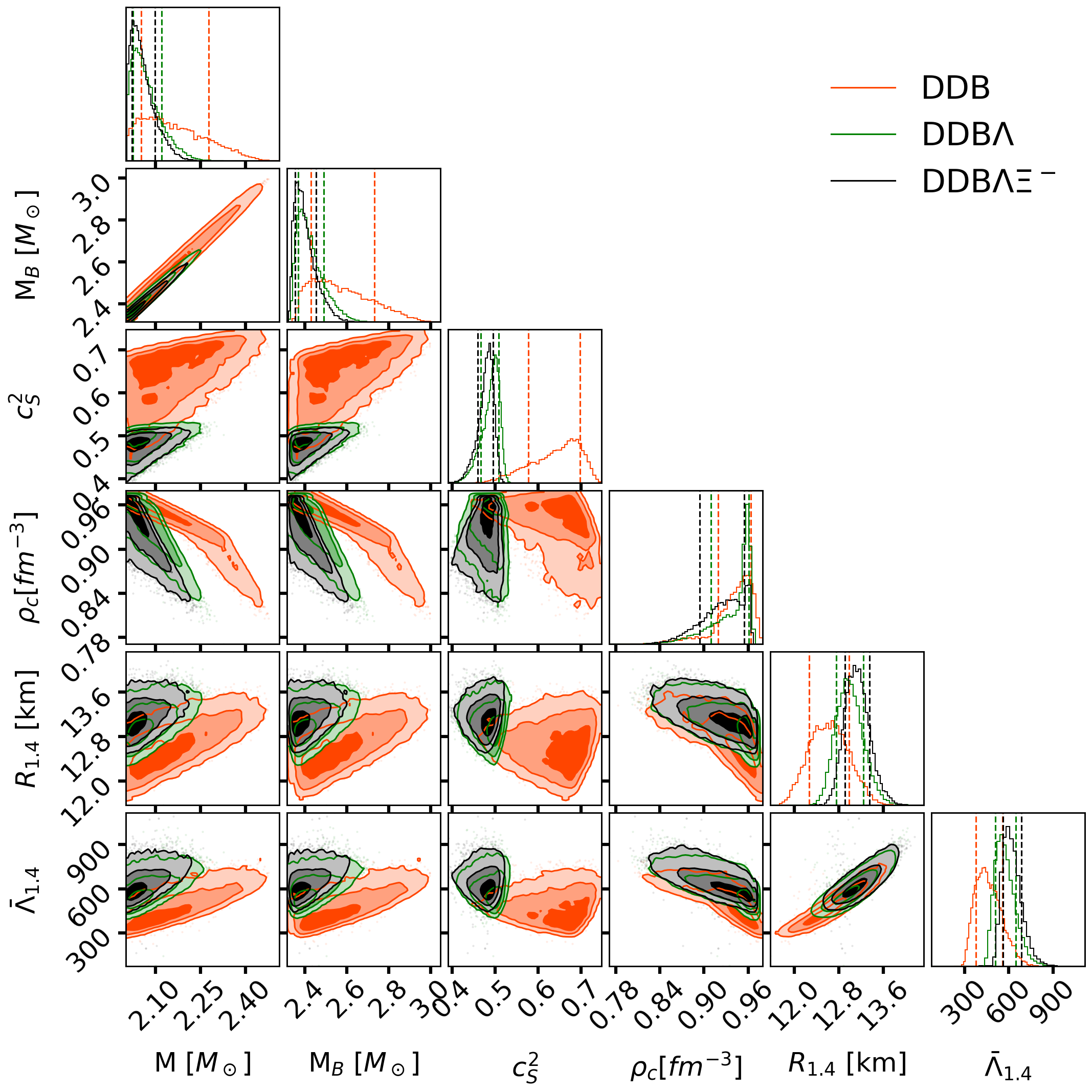}
\caption{Corner plot for the  marginalized posterior distributions of some  NS properties:  gravitational mass $M_{\rm max}$, baryonic mass $M_{\rm B, max}$, the square of central speed-of-sound $c_s^2$, the central baryonic density $\rho_{c}$, the radius $R_{1.4}$ and  the dimensionless tidal deformability $\bar \Lambda_{1.4}$ for 1.4 $M_\odot$ NS
for the DDB (red), DDB with $\Lambda$ hyperon (green) and DDB with $\Lambda$ and $\Xi^-$ hyperon (black)  sets. 
The vertical lines indicate the  68\%  CI,  and the dark to light intensities represent, respectively, the 1$\sigma$, 2$\sigma$, and 3$\sigma$ CI.  \label{T:fig3b}}
\end{figure*}
{Some conclusions can be drawn from the above PDs of NMPs and NS properties: the joint inclusion of hyperons and  the two solar mass constraint has major effects on the iso-scalar channel. For example, the median values of nuclear incompressibility $K_0$ are  231 MeV, 260 MeV and  273 MeV respectively, for DDB, DDB$\Lambda$ and DDB$\Lambda\Xi^-$. It is to be noted that $K_0=230\pm40$ MeV is in part of the fit data for all the three cases, however the lower values of $K_0$ are not able to produce 2 M$_\odot$ NS in the  presence of $\Lambda$ and $\Lambda-\Xi^-$ hyperons in the  DDB$\Lambda$ and DDB$\Lambda\Xi^-$ sets, respectively.  A similar behavior can also be seen for the skewness coefficient $Q_0$. On the other hand, the NMPs patterning to the density dependence of the symmetry energy remain similar in all the three cases. Only minor changes ($\sim 1-2\%$) are observed in the curvature parameter of the symmetry energy $K_{\rm sym,0}$. 

Next we  analyze the NS properties of the three EOSs sets. The hyperons have  a major impact on the star properties. The 90\% CI maximum limits of NS maximum mass reduces by 8\% and 9\% in case of the DDB$\Lambda$ and DDB$\Lambda\Xi^-$ sets compared to the DDB set, respectively. This arises because the presence of hyperons disfavors large maximum masses.
As a consequence, the tail of the maximum mass distribution of the DDB set that extends beyond 2.4$M_\odot$, stops at $\approx 2.25M_\odot$ for the other two sets. The well defined peaks of the DDB$\Lambda$ and DDB$\Lambda\Xi^-$ distributions are due to the fact that the three sets have the same number of EOS, but the two sets that include hyperons have all the maximum  mass stars in a smaller interval, $2\lesssim M_{\rm max}\lesssim2.25M_\odot$.

The clear distinction between only nucleonic and  nucleonic-hyperonic sets is also seen in Fig. \ref{T:fig3b} where the distribution of the square of the speed-of-sound at the NS centre, $c_s^2(\rho_c)$, is plotted.   Hyperons give rise to a strong reduction of the speed of sound in the centre of the star, being larger if the two hyperons are included. This is a reflection of the softening of the EOS due to the onset of new degrees of freedom.

The NS radius and the tidal deformability for 1.4 M$_\odot$ stars also show noticeable changes in the hyperonic sets compared with the nucleonic set. It was discussed above that the presence of hyperons together with the two solar mass constraint give rise to  stiffer iso-scalar EOS. A direct consequence is seen in the larger radii and tidal deformabilities predicted by the sets DDB$\Lambda$ and DDB$\Lambda\Xi^-$ for radius and tidal deformability of the 1.4$M_\odot$ stars.

Some other NS properties are given in Table \ref{tab4}, in particular, the median, maximum and minimum of the 90\% CI of the radii and tidal deformabilities of stars with masses 1.6, 1.8, 2.075 and 2.2 $M_\odot$ as well as the effective tidal deformability $\tilde\Lambda_{q=1}$ of a NS binary of two equal mass stars with a chirp mass 1.186$M_{\odot}$, $q$ is the mass ratio of NSs involved in binary merger. The trend on the radii is the same seen in the corner plot Fig.  \ref{T:fig3a}: the two solar mass constraint and the presence of hyperons results in larger  0.15 to 0.3 km maximum limits and 0.7 to 1 km minimum limits, and similar effects on the tidal deformabilities. The effective tidal deformability $\tilde\Lambda_{q=1}$ is compatible with results from the GW170817 detection \cite{LIGOScientific:2018hze}. We also include in the table information on the hyperon fraction in the center of maximum mass stars. If only $\Lambda$s are included the hyperon fraction take values above  45\%. In a calculation with $\Lambda$s and $\Xi^-$s the hyperon is just slightly smaller, but above 40\%.

{
In order to better understand the model DDH used in the present study and the consequences of introducing hyperons subject to the two solar mass constraint,  we show in Fig. \ref{M:fig2a} the
conditional probabilities distribution $P(R|M)$ for the nucleonic EOSs corresponding to the three sets  DDB, DDB$\Lambda$ and DDB$\Lambda\Xi^-$ ("no hyperons" are include in the last two sets). It is to be noted that the "no hyperons" refers to the case where we obtain nucleonic EOS with the parameter distribution of hyperonic sets by switching off the hyperons. The left  panel of  Fig. \ref{M:fig2a} shows that the DDH model and the parameter domain defined in Table \ref{tab2} spans the whole space of interest. To exemplify this statement we plot on top  of the 99\% CI  M-R curves for the nucleonic EOSs within DDH (DDB, DDB$\Lambda$ and DDB$\Lambda\Xi^-$) sets  the M-R curves obtained within several RMF models with average and extreme properties:  DD2 \cite{Typel2009}, DDME2 \cite{Lalazissis2005}, DDMEX and  DDLZ1 \cite{Huang:2020cab}, FSU2R \cite{Tolos:2017lgv}, SFHo \cite{Steiner:2012rk},   TW \cite{Typel1999}. Our parametrization  at 99\% CI covers almost all the RMF EOS. In the same figure,  the pink band  represents a large set of Skyrme forces \cite{Malik:2022spt}, which can describe stars with smaller radii but are not intrinsically causal, and it was an objective of the present study to work within a framework that by construction includes causality. In the right  panel of  Fig. \ref{M:fig2a}, we show the 90\% CI M-R bands obtained for the nucleonic DDB (grey band), DDB$\Lambda$ (dotted band) and DDB$\Lambda\Xi^-$ (slashed band)  sets, i.e. no hyperons are included in the last two sets.
The main conclusion is that if hyperons really exist inside NS that attain 2$M_\odot$ then the nuclear matter properties correspond to stiffer EOS, that give rise to larger radii and  slightly larger maximum masses, a similar conclusion was drawn in \cite{Fortin:2014mya}.   The sets of nucleonic EOS corresponding to the DDB$\Lambda$ and DDB$\Lambda\Xi^-$ sets could also have been generated within the DDB models but would lie outside the  DDB 90\% CI region.}

\begin{figure*}
\begin{tabular}{cc}
\includegraphics[width=0.5\textwidth]{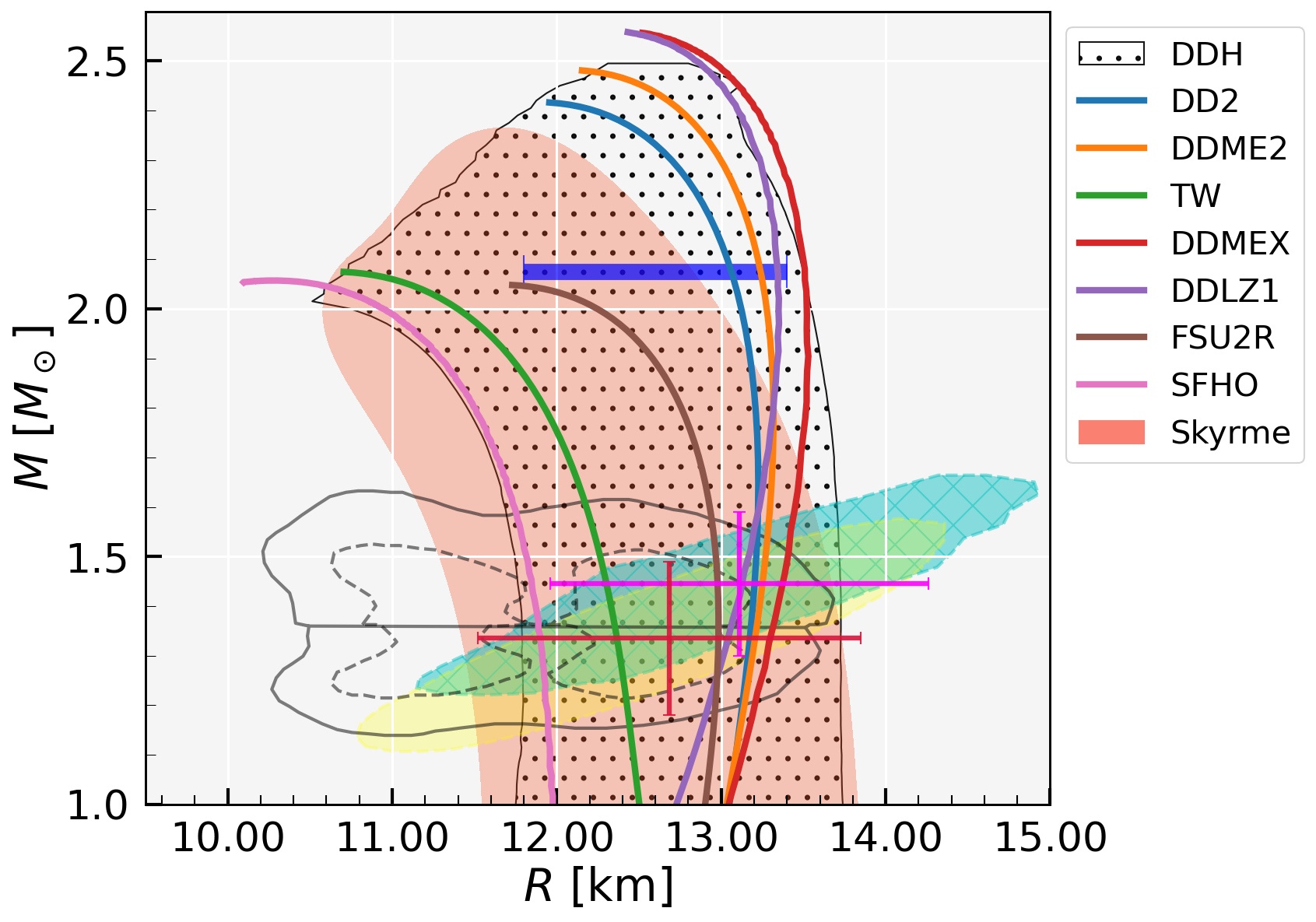}      &
\includegraphics[width=0.42\textwidth]{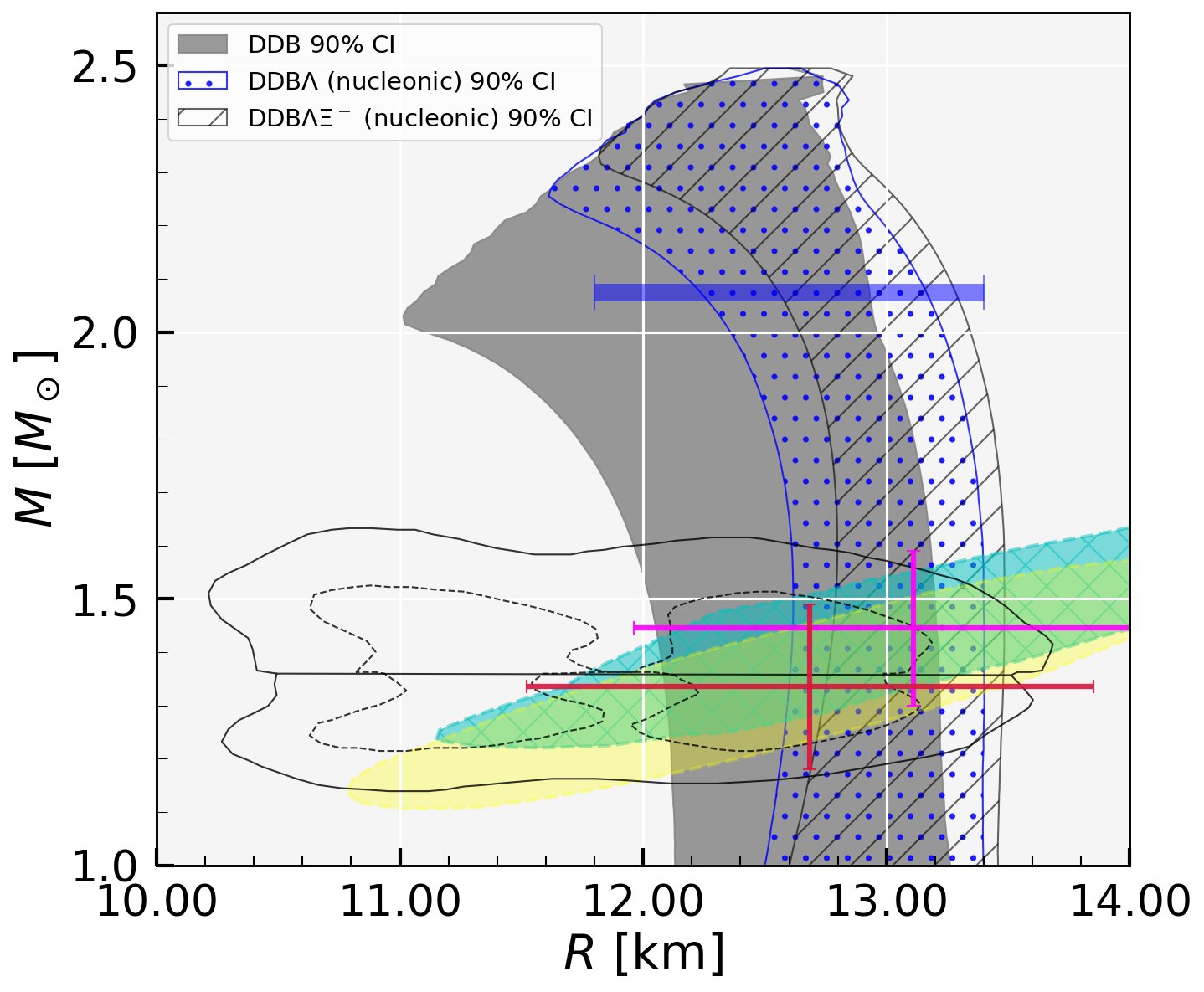} \\ 
\end{tabular}
\caption{{Left panel:  99\% CI of the conditional probabilities distribution $P(R|M)$ obtained for the nucleonic DDH EOSs with DDB, DDB$\Lambda$ and DDB$\Lambda\Xi^-$  sets  (dotted region) together with the M-R curves determined for some RMF models with  average and extreme properties: DD2 \cite{Typel2009}, DDME2 \cite{Lalazissis2005}, DDMEX and  DDLZ1 \cite{Huang:2020cab}, FSU2R \cite{Tolos:2017lgv}, SFHo \cite{Steiner:2012rk},   TW \cite{Typel1999}. The pink band represents a set of  MR curves obtained in the framework of Skyrme forces \cite{Malik:2022spt}.
Right panel: conditional probabilities $P(R|M)$ region defined by the nucleonic DDB set (grey region), the nucleonic DDB$\Lambda$ set (dotted region) and the nucleonic DDB$\Lambda\Xi^-$ set (slashed region) at 90\% CI.
In both panels, the 90\% (solid) and 50\% (dashed) CI  for the binary components of the  GW170817 event  \cite{LIGOScientific:2018hze} are represented by the gray regions in the left panel. The $1\sigma$ (68\%) confidence region for the 2-D posterior distribution in mass-radii domain from the millisecond pulsar PSR J0030+0451 NICER x-ray data (the cyan hatched and yellow) \cite{Riley:2019yda,Miller:2019cac}. The horizontal (radius) and vertical (mass) error bars represent the $1\sigma$ confidence interval obtained for the 1-D marginalized posterior distribution  of the same NICER data. The blue bars in both panel represents the 90\% CI radius of the PSR J0740+6620 with 2.08$M_\odot$  \cite{Miller:2021qha}. 
}
\label{M:fig2a}} 
\end{figure*}

\begin{figure*}
\includegraphics[width=0.98\textwidth]{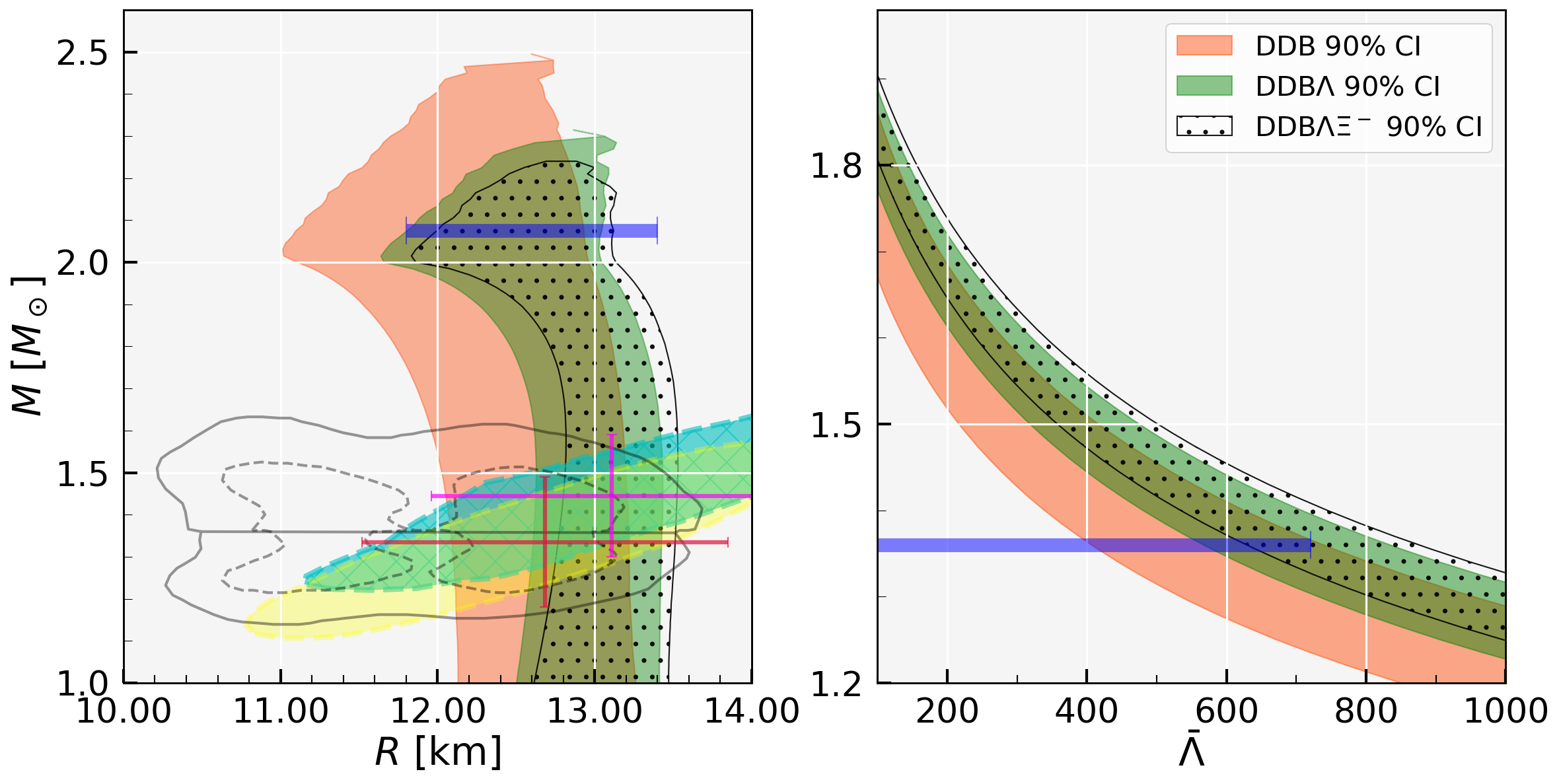}
\caption{The 90\% CI region calculated from the conditional probabilities $P(R|M)$ (left) and $P(\Lambda|M)$ (right) for the sets DDB  (red), DDB$\Lambda$ (green) and DDB$\Lambda\Xi^-$ (dotted). The 90\% (solid) and 50\% (dashed) CI  for the binary components of the  GW170817 event  \cite{LIGOScientific:2018hze} are represented by the gray regions in the left panel. {The $1\sigma$ (68\%) confidence region for the 2-D posterior distribution in mass-radii domain from the millisecond pulsar PSR J0030+0451 NICER x-ray data (the cyan hatched and yellow) \cite{Riley:2019yda,Miller:2019cac}. The horizontal (radius) and vertical (mass) error bars represent the $1\sigma$ confidence interval obtained for the 1-D marginalized posterior distribution  of the same NICER data.}
The blue bars represent:  the 90\% CI radius of the PSR J0740+6620 with 2.08$M_\odot$  \cite{Miller:2021qha} (left panel) and the 90\% CI   tidal deformability of a 1.36$M_\odot$ star \cite{LIGOScientific:2018cki} (right panel).
\label{M:fig2}} 
\end{figure*}

{In Fig. \ref{M:fig2}, we plot the 90\% CI region obtained from the conditional probabilities $P(R|M)$ (left) and $P(\Lambda|M)$ (right) for the posterior distributions of the DDB, DDB$\Lambda$ and DDB$\Lambda\Xi^-$, i.e. for a given mass $M$ the 90\% CI distribution of the radius $R$ and the tidal deformability $\bar \Lambda$ is shown. We represent observational constraints with two blue bars: 
the 90\% CI radius for a 2.08$M_\odot$ star determined in \cite{Miller:2021qha} is included in the left panel and  the  90\% CI determined in \cite{LIGOScientific:2018cki} for the 1.36$M_\odot$ tidal deformability in the right panel.
The gray regions indicate the 90\% (solid) and 50\% (dashed) CI associated with the  binary component of the  GW170817 event \cite{LIGOScientific:2018hze}.
We have also included the constraints set by the PSR J0030+0451 NICER x-ray data  \cite{Riley:2019yda,Miller:2019cac}. These figures confirm the trend already discussed: NS with a mass above 2 M$_\odot$ has larger radii and subsequently larger tidal deformations if  hyperons are included. The present observational constraints either from LIGO-Virgo or from NICER can not rule out the presence of hyperons in the core of a NS. We conclude that future precise measurement of simultaneous mass and radius or tidal deformation of NS masses above 2 M$_\odot$ maximum mass will bring some information on the possible presence of hyperonic degrees of freedom inside NS core: the measurement of a  two solar mass star with $R<11.5$ km would indicate that no hyperons are present in the NS core.}

\begin{figure}
\includegraphics[width=0.45\textwidth]{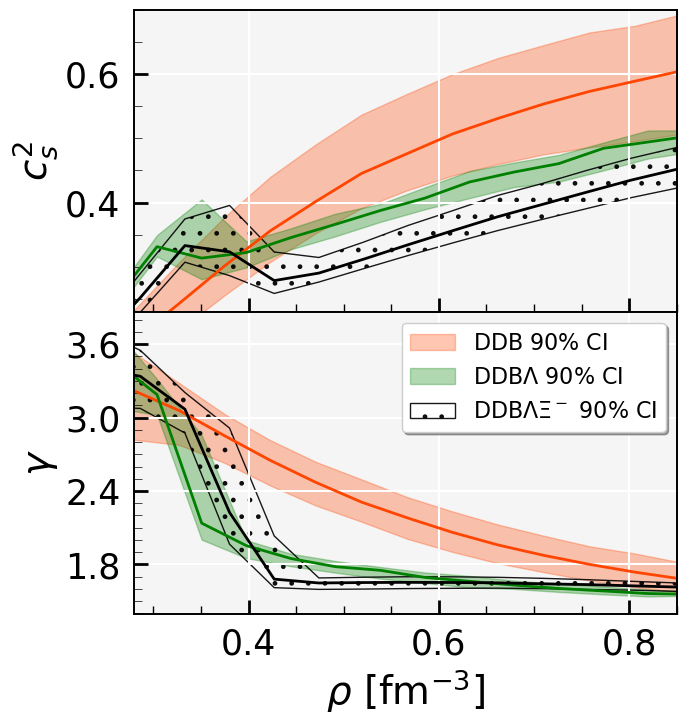}
\caption{The median and 90\% CI of the square of sound velocity $c_s^2$ (top) and the polytropic index $\gamma$ introduced in \cite{Annala:2019puf} (bottom) as a function of baryon density for DDB, DDB$\Lambda$ and DDB$\Lambda\Xi^-$ sets.} 
\label{T:cs}
\end{figure}

{We have already referred to the effect of hyperons on the speed of sound at the centre of the NS. It is, however, interesting to discuss the behavior of the speed of sound with density within the three sets DDB, DDB$\Lambda$ and DDB$\Lambda\Xi^-$, see Fig. \ref{T:cs} (top). It was shown that  2$M_\odot$ NSs requires a speed of sound well above the conformal limit $1/\sqrt{3}$ and which implies that neutron star matter is a strongly interacting system \cite{Bedaque:2014sqa,Alford:2013aca,Moustakidis:2016sab,Tews:2018kmu,Reed:2019ezm}. 
Compared with the DDB set, the appearance of $\Lambda$ and $\Lambda-\Xi^-$ hyperons gives rise to: an increase of the speed of sound below $\sim 2\rho_0$, followed by a  decrease of $c_s^2$ between 2 and 3$\rho_0$ and finally a  soft increase of  $c_s^2$ until values below 0.5 $c^2$ at the center of the star, well below the $\sim 0.7c^2$ obtained with DDB.}

{Finally, let us also discuss the polytropic index introduced in \cite{Annala:2019puf} $\gamma= \partial\ln p/\partial\ln\epsilon$ as a quantity that could convey information on a possible phase transition to deconfined quark matter. In this study, where a huge set of EOS using a speed of sound interpolation method connecting the low densities constrained by the $\chi$EFT neutron matter pressure and the very high densities  constrained by to pQCD,  it was suggested that $\gamma\approx1.75$ could be identified with the onset of quark matter. The authors also indicate  $\gamma\approx2.5$ as characterizing  the EOS at saturation or just above. In Fig. \ref{T:cs} (bottom), the $\gamma$ index is plotted for the three sets of models used in the present study, for densities above the saturation density. We confirm that, indeed, for densities just above  2$\rho_0$ the $\gamma$ takes values of the order of 2.4-3.0. With the onset of hyperons the $\gamma$ index reduces quite steeply and takes values of the order of $\approx 1.6 - 1.8$ above 3$\rho_0$. On the other hand, the $\gamma$ index associated with the EOS of the nucleonic set DDH decreases smoothly and attains the value $\approx$1.75 above 5$\rho_0$. These results seem to indicate that although a $\gamma$ index of the order of 1.75 at not too large densities may indicate the onset of some kind of exotic matter, it is not necessarily associated with quark matter. Nucleonic matter may also attain a value of the order of  $\approx$1.75 but for much larger densities.
}

\section{Summary and Conclusions \label{sec:con}}
We have analyzed  the possible signatures of the presence of hyperons inside NS. The study was undertaken within a relativistic mean field description of hadronic matter considering density dependent couplings \cite{Typel1999,Malik:2022zol}. {Three sets ($\sim 15,000$ EOSs in each set) were generated using a Bayesian inference approach and imposing a minimal set of nuclear matter constraints  and a  NS  maximum mass  above 2M$_\odot$: DDB set including just nucleons, DDB$\Lambda$  with nucleons and $\Lambda$s and DDB$\Lambda\Xi^-$  with nucleons, $\Lambda$s and $\Xi^-$.} The hyperon couplings were fixed taking into account the present information on hypernuclei \cite{Chatterjee:2015pua,Gal:2016boi,Fortin:2017cvt,Fortin:2017dsj,Providencia:2018ywl}, and considering the SU(6) quark model predictions for the isoscalar vector meson couplings. All the three sets were constrained to describe a two solar mass NS.

We could conclude that the 2$M_\odot$ constraint has strong effects on the NS properties if hyperonic degrees of freedom are included.  This is because the joint two solar mass constraint and the onset of new degrees of freedom imposes a harder isoscalar channel with larger isoscalar NMP as the $K_0$, $Q_0$ and $Z_0$. In particular, the incompressibility $K_0$ is in average 30 to 50 MeV larger than the one obtained for the DDH set, however, still taking values that are perfectly within the accepted ones. One the other hand, the isosvector channel is not affected by the inclusion of hyperons. The harder EOS have noticeable effects on the star radius, giving rise to larger radii, both for stars with a mass close to the canonical  one or with a mass around two solar masses.  

Some observations that would exclude the presence of hyperons, or reduce it to a negligible fraction, would be: the detection of a 1.4$M_\odot$ star with a radius below 12.5 km; a two solar mass star with a radius below 11.5 km; or a maximum mass above $\sim 2.2$ $M_\odot$.

The results of the present study corroborate and generalize the conclusions drawn in \cite{Fortin:2014mya}, where  the properties of NS obtained with 14  EOS of dense matter with hyperons,  which predict NS with at least two solar masses, were  compared with the properties of three widely used nucleonic EOS.  In this study, the authors refer that the 14 hyperonic EOS used in their analysis do not satisfy the PNM constraints on the pressure obtained  in the chiral EFT calculation of \cite{Hebeler2013}. They conclude that   models that predict a sizable amount of hyperons in massive NS also predict for  $1.0-1.6\, M_\odot$ NS radii  above  13 km. On the other hand, if these stars have radii below 12 km, sizable hyperon cores are ruled out in massive stars. Notice, however, that in \cite{Fortin2016} it was shown that some of the hyperonic EOS  studied were predicting radii below 13 km for 1.4$M_\odot$ stars. These were the DD2 \cite{Typel2009} and DDME2  \cite{Lalazissis2005} models. In the present study,  the  three sets of EOS  used to draw our conclusions all satisfy the PNM constraints of \cite{Hebeler2013} within 2$\sigma$, also the sets DDB$\Lambda$ and DDB$\Lambda\Xi^-$. It is true, however, that the hyperonic sets show a stiffer behavior than the nucleonic set.  We also conclude that if the two solar mass maximum mass constraint is imposed,  NS from the DDB$\Lambda$ and DDB$\Lambda\Xi^-$ have in average larger radii. In particular,  1.4$M_\odot$ NS with radii $\lesssim12$ km were only obtained with the DDB set, although stars with $R_{1.4}\gtrsim 12.5\,$ km are present in the   DDB$\Lambda$ and DDB$\Lambda\Xi^-$ sets.

{Finally, it is shown that  if the  polytropic $\gamma$ index,  introduced in \cite{Annala:2019puf}, takes values of the order of 1.75 at not too large densities, $\approx3\rho_0$, it may indicate the onset of some kind of exotic matter, which, however, is not necessarily quark matter. In the present study, it is hyperonic matter. The $\gamma$ index of  nucleonic matter may also attain values $\approx$1.75 but at much larger densities.
}

\begin{acknowledgements}
This work was partially supported by national funds from FCT (Fundação para a Ciência e a Tecnologia, I.P, Portugal) under the Projects No. UID/\-FIS/\-04564/\-2019, No. UIDP/\-04564/\-2020, No. UIDB/\-04564/\-2020, and No. POCI-01-0145-FEDER-029912 with financial support from Science, Technology and Innovation, in its FEDER component, and by the FCT/MCTES budget through national funds (OE). The authors acknowledge the Laboratory for Advanced Computing at University of Coimbra for providing {HPC} resources that have contributed to the research results reported within this paper, URL: \hyperlink{https://www.uc.pt/lca}{https://www.uc.pt/lca}.
\end{acknowledgements}

%
\end{document}